\documentclass[conference]{IEEEtran}
\IEEEoverridecommandlockouts

\usepackage{cite}
\usepackage{amsmath,amssymb,amsfonts}
\usepackage{algorithmic}
\usepackage{graphicx}
\usepackage{textcomp}
\usepackage{xcolor}
\def\BibTeX{{\rm B\kern-.05em{\sc i\kern-.025em b}\kern-.08em
 T\kern-.1667em\lower.7ex\hbox{E}\kern-.125emX}}
 
\usepackage{csquotes}
\usepackage{caption}
\usepackage{subcaption}
\usepackage{graphicx}
\usepackage{listings}
\usepackage{tikz}
\usepackage{todonotes}
\usepackage{hyperref}

\newcommand*\circled[1]{\tikz[baseline=(char.base)]{\node[shape=circle,draw,inner sep=1pt] (char) {#1};}}
\definecolor{backcolour}{rgb}{0.98, 0.98, 0.97}

\begin{document}

\title{Gromit: Benchmarking the Performance and Scalability of Blockchain Systems\thanks{This work was funded by NWO/TKI grant BLOCK.2019.004}}

\author{
 \IEEEauthorblockN{Bulat Nasrulin, Martijn De Vos, Georgy Ishmaev, Johan Pouwelse}
 \IEEEauthorblockA{Delft University of Technology
 \\\{b.nasrulin, m.a.devos-1, g.ishmaev, j.a.pouwelse,\}@tudelft.nl}
}
\maketitle

\begin{abstract}
	The growing number of implementations of blockchain systems stands in stark contrast with still limited research on a systematic comparison of performance characteristics of these solutions. Such research is crucial for evaluating fundamental trade-offs introduced by novel consensus protocols and their implementations. These performance limitations are commonly analyzed with ad-hoc benchmarking frameworks focused on the consensus algorithm of blockchain systems. However, comparative evaluations of design choices require macro-benchmarks for uniform and comprehensive performance evaluations of blockchains at the system level rather than performance metrics of isolated components. 
	To address this research gap, we implement Gromit, a generic framework for analyzing blockchain systems.
	Gromit treats each system under test as a transaction fabric where clients issue transactions to validators.
	We use Gromit to conduct the largest blockchain study to date, involving seven representative systems with varying consensus models.
	We determine the peak performance of these systems with a synthetic workload in terms of transaction throughput and scalability and show that transaction throughput does not scale with the number of validators. We explore how robust the subjected systems are against network delays and reveal that the performance of permissoned blockchain is highly sensitive to network conditions. 
\end{abstract}


\section{Introduction}

The rapid growth in the number of blockchain protocols in the past few years has been boosted by the interest in crypto-currencies, decentralized finance, and identity systems. The solutions are mostly empirically driven, with direct economic incentives stimulating engineering experiments. To date, there are more than 700 different blockchain and distributed ledger platforms offering native digital assets and products.
Many of these solutions deploy original families of consensus protocols or significant modifications of popular protocols, with variations in scalability, performance, and decentralization guarantees. These developments outpace systematization and research on inherent trade-offs of different design choices~\cite{fan2020performance}. 

The absence of benchmarking solutions for comprehensive comparative analysis of various protocols is a particularly problematic omission, given the cumulative marketcap of 1,468 trillion \$ for these projects. More fundamentally, this systematization gap hampers our ability to tackle the increasing complexity of blockchain systems and make conscious design choices in blockchain engineering. 
Developers of these protocols often provide performance metrics of blockchain solutions as declarative whitepapers that do not pass the standards of peer review and reproducibility, calling into question the reliability and objectivity of these measurements. For instance, it is a common practice to provide performance metrics of an isolated component and report them as a system-wide performance metric~\cite{rocket2019scalable, lokhava2019fast, cao2020performance}. This practice often leads to false impressions of the end-to-end system performance.

Few available benchmarking solutions, such as Blockbench~\cite{dinh2017blockbench} and Hyperledger Caliper~\cite{caliper} focus on narrow sets of permissioned consensus protocols or DAG-based protocols as DAGBENCH~\cite{dong2019dagbench}. There is also a noticeable deficit of academic research in macro benchmarks for blockchain systems. Existing studies are rather limited in scope either focusing on specific protocols such as Hyperledger~\cite{sukhwani_2018, kuzlu_2019} or Ripple~\cite{touloupou_2021}, or reusing Hyperldeger Caliper~\cite{dabbagh_2020} and Blockbench~\cite{dinh2017blockbench}. BCTMark is one of the few comparative benchmark studies which compares three different protocols, including permissionless Ethereum blockchain~\cite{saingre2020bctmark}. The authors in this study highlight the necessity to extend the comparison set and include more metrics such as partition tolerance. The most recent systematic survey on performance evaluation of blockchain systems demonstrates that available comparative studies are rather limited in scope both in terms of compared systems and depth of analysis, focusing on isolated layers of blockchain systems~\cite{fan2020performance}. To address this research gap, we design a benchmark that is comprehensive in scope, allowing us to stress-test the system under different network conditions.

\begin{figure*}[t]
	\center
	\includegraphics[width=0.88\linewidth]{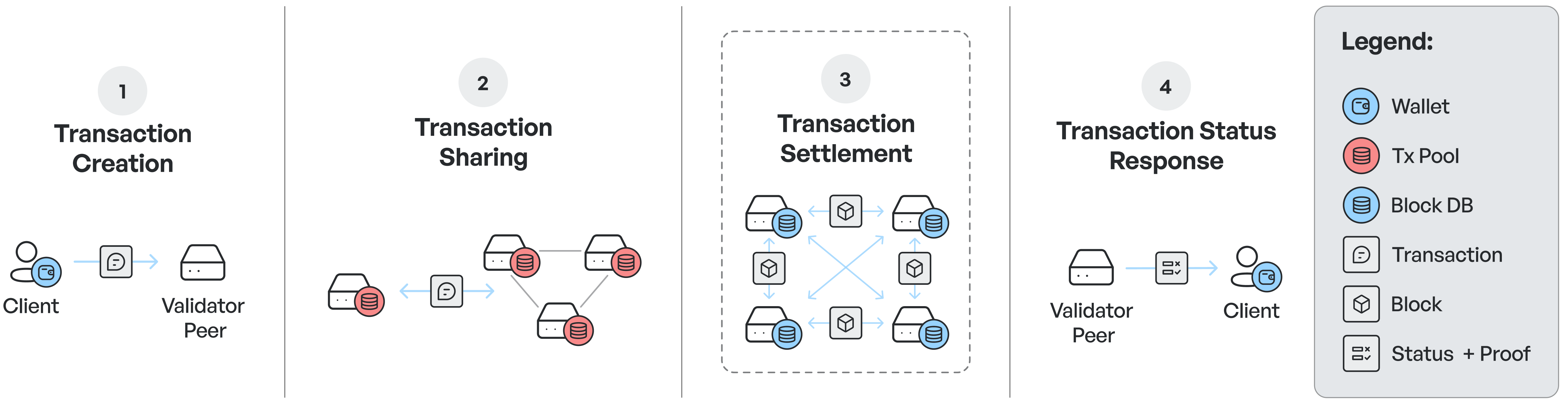}
	\caption{Our abstraction model of blockchain system as a transaction fabric, comprising four stages. The dotted line highlights the focus of most whitepaper benchmarks.}
	\label{fig:blockchain_model}
\end{figure*}

We introduce \emph{Gromit}, a generic benchmarking framework that allows a performance evaluation of \emph{any} blockchain solution.
Gromit treats each system under test as a transaction fabric, which means a transaction processing system where a group of peers continuously reach a consensus on transactions submitted by clients. Gromit analyses performance metrics related to transactional data, specifically throughput and latency.
As we will show, this data alone can reveal the limitations of various aspects of the system.

We show the applicability of Gromit{} and conduct the \emph{largest blockchain benchmarking study to date}.
Our benchmark involves seven major blockchain solutions with different consensus algorithms. We determine the peak performance of these blockchain systems for different numbers of peers and without any modifications to the source code.
A key finding is that the performance for most evaluated blockchain systems \emph{degrades} when the number of peers increases.
We also reveal the effect of network delays and an increase in system load on the distribution of end-to-end transaction latencies, yielding valuable insights into the real-life bottlenecks of underlying consensus mechanisms. While most performance bottleneck research focuses on the consensus layer, a system-wide stress test can reveal bottlenecks in other layers, e.g., in the persistence layer.

The contribution of this work is two-fold:
\begin{enumerate}
	\item We design and implement \emph{Gromit}, a benchmarking framework that enables an analysis of \emph{any} blockchain system (Section~\ref{sec:Prickle_architecture}).
	\item We conduct \emph{the largest blockchain benchmark to date}, measuring the performance of seven prominent blockchain systems (Section~\ref{sec:evaluation}).
\end{enumerate}








\section{Blockchain as a Transaction Fabric}
\label{sec:blockchain_as_tx_fabric}

In this work, we view a blockchain system as a \emph{transaction fabric}.
Figure~\ref{fig:blockchain_model} visualizes this abstraction model, which illustrates a typical transaction life cycle. 

In our model, we distinguish between clients and peers. Clients are instances that create transactions and send them to peers. An example of a client is a light wallet or lightweight nodes in a Bitcoin network. Peers in our model are responsible for processing and validating transactions in a shared, decentralized network. Thus, we call them \emph{validator peers}. In some blockchain systems, they are also referred to as miners.

\subsection{Transaction Life Cycle Model}

\subsubsection{Transaction Creation}
A \textit{transaction} contains logic that modifies the system state, e.g., by transferring an asset to another account. Each transaction is cryptographically signed with the private key of the issuing client to ensure authenticity. 
Blockchain solutions usually provide \textit{Wallet} API's for clients to submit their transactions, e.g., with an RPC endpoint or REST endpoint. 

\subsubsection{Transaction Sharing}
Blockchain systems employ complex transaction sharing mechanisms. Permissionless blockchains typically use a global gossip protocol to share transactions over a structured or unstructured overlay. Permissoned blockchains are deployed in a more controlled network environment and, as a result, might share transactions using a broadcast algorithm. 

The transactions are stored in a datastore, often called a transaction pool (\textit{TxPool}). The transaction pool is a temporary store used to queue or preprocess transactions before the network validates them. 

\begin{figure*}[t]
	\centering
	\includegraphics[width=0.80\linewidth]{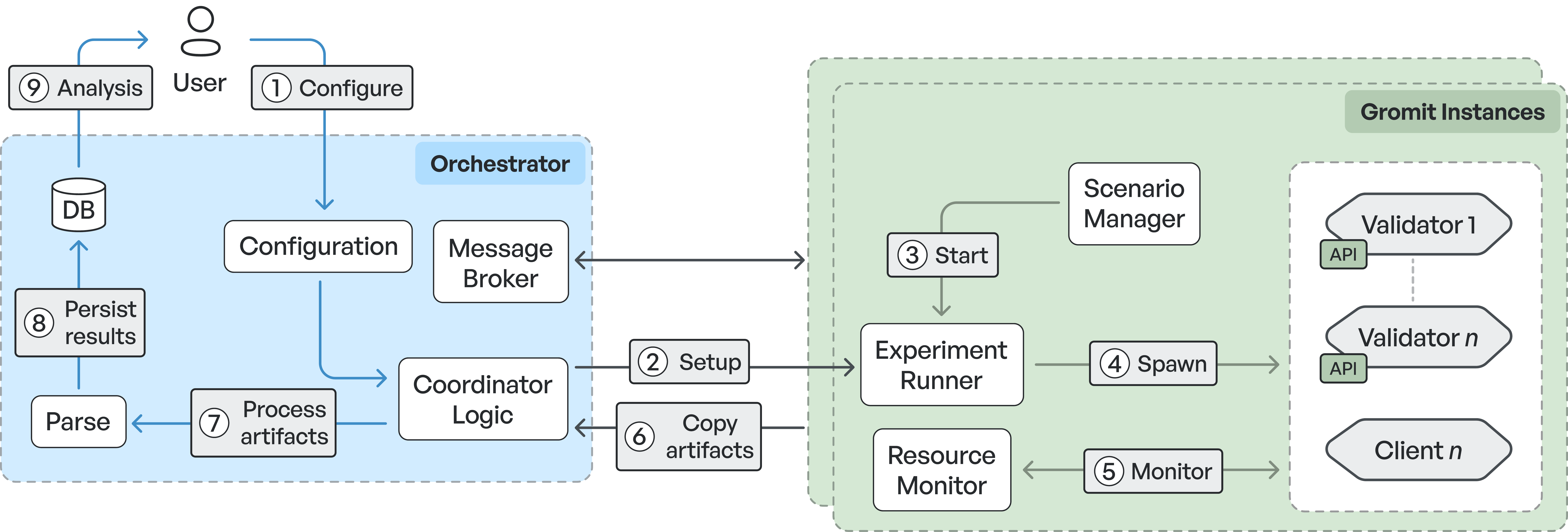}
	\caption{Architectural overview and process flow of Gromit, our benchmarking framework for blockchain systems.}
	\label{fig:Prickle_architecture}
\end{figure*}

\subsubsection{Transaction Settlement}
A consensus algorithm is a crucial part of the blockchain system as a mechanism for achieving security and liveness\cite{bano2019sok}. 
The intermediate outcome of the consensus process in blockchain systems is a set of valid transactions. Valid transactions are stored in a tamper-proof distributed ledger, a replicated data structure maintained by validators. The system discards invalid transactions.

Blockchain solutions typically bundle valid transactions in \textit{blocks}, interlinked in a hash chain, and stored in a local database (\textit{Block DB}). 
Each block in a hash chain contains the cryptographic hash of the previous block, making illegitimate modifications of the chain detectable. Some blockchain-like systems adopt a different organization of the distributed ledger, e.g., by maintaining a Directed Acyclic Graph (DAG)~\cite{dong2019dagbench}. 

\subsubsection{Transaction Status Response}
After the transaction is settled the client waits for a transaction approval (or rejection), received from validator peers. 
Some blockchains also include a \textit{proof} in the response that proves that a transaction is finalized. 

\subsection{Transaction Performance Indicators}

Our approach is to analyze the performance of current blockchain solutions through \emph{transaction benchmarking}.
The speed at which a blockchain system processes transactions is a defining metric for blockchains.
High transaction latencies directly impact end-users experience, e.g., the relatively high finalization times of Bitcoin transactions (10 minutes) make it unsuitable for interactive trade~\cite{bamert2013have}.
We include all stages of the transaction life cycle in our measurements, rather than focusing on the performance of the consensus layer only.

We obtain insights into the limitation of blockchain systems by measuring peak performance and associated transaction latencies. 
We also measure performance under different network conditions, such as changes in the geographical distribution of the overlay network. Our experiments in Section~\ref{sec:evaluation} highlight that these metrics can reveal performance bottlenecks and act as a guideline for optimization efforts.

\section{Gromit: A Generic Framework for Blockchain Benchmarking}
\label{sec:Prickle_architecture}
We design and implement \emph{Gromit}, the first generic framework for blockchain benchmarking.
Gromit is designed for quick experiment iterations, utilizing a domain-specific language to devise experiments.
Our framework, implemented in Python, is based on the abstraction model discussed in Section~\ref{sec:blockchain_as_tx_fabric}.
To encourage the adoption of Gromit, its source code is published on Github.\footnote{See \url{https://github.com/grimadas/gromit}}
Gromit provides developers and researchers with the necessary tools to design benchmarks targeting blockchain performance.
Gromit spawns client and validator processes on remote servers and coordinates interactions between running processes, e.g., transaction issuing.

\subsection{Experiment Flow}
Figure~\ref{fig:Prickle_architecture} shows the architecture of the framework and the flow of an experiment.
This diagram is described next.

\textbf{Experiment Setup.}
Before an experiment starts, it spawns a dedicated process, called the \emph{orchestrator}, that setups the environment on remote servers and handles cross-server communication.
The orchestrator reads the configuration file associated with the experiment.
This file specifies the network addresses of remote servers and the number of clients and validators that should be started (step~\circled{1}).
The orchestrator then copies the source code and all necessary experiment files to the remote servers using \texttt{rsync}. 
Next, the coordinator setups the environment on the specified servers over an SSH connection (step~\circled{2}).
This step includes the installation of required system packages and the generation of a genesis file.
This file describes the initial state of the blockchain system and can pre-load specific accounts with assets. 
The orchestrator then starts a \emph{instance} for each client or validator on the remote server and assigns an identifier to each running instance.
The scenario manager, part of the logic of a Gromit instance, parses a provided scenario file and starts the experiment (step~\circled{3}).

\textbf{Scenario Files.}
A user describes the actions performed during an experiment with a \emph{scenario file}.
The scenario manager parses this file and schedules actions using the \texttt{asyncio} library.

\begin{lstlisting}[basicstyle=\small,caption={An example of a scenario file in Gromit, describing an experiment with 10 clients and validators.},captionpos=b,label={lst:scenario},frame=single]
@8   init_blockchain_config {1-10}
@12  start_validator {1-10}
@20  start_client {11-20}
@80  start_creating_transactions {11-20}
@100 stop
\end{lstlisting}

An example scenario file associated with a simple blockchain benchmark is given in listing~\ref{lst:scenario}.
Actions and their timestamps are explicitly denoted.
A user can schedule an action to execute on a subset of all validators.
Scenario files can be machine-generated and are a flexible approach to devising and conducting experiments.

\textbf{Experiment Runner.}
The experiment runner spawns validators and clients as a subprocess during each experiment run (step~\circled{4}).
Clients interact with validators through the exposed API endpoint.
We simulate a client in Gromit as a procedure that issues transactions.
During an experiment, Gromit instances can share data over a TCP connection through a message broker.
We use the message broker functionality to share the credentials of pre-defined blockchain accounts with clients.
Gromit contains tools to gracefully terminate a running experiment, e.g., when a particular condition is not met.

Gromit also provides utilities to track system resource usage.
Gromit instances monitor CPU, memory, disk, and network usage using the \texttt{procfs} library (step~\circled{5}).
These metrics allow us to estimate the system resource usage of blockchain systems.
Developers can easily extend Gromit to monitor specific metrics, for example, the number of inbound network messages for a particular blockchain system.

\textbf{Collecting Experiment Results.}
When the experiment ends, the orchestrator copies all generated artifacts from the remote nodes using \texttt{rsync} (step~\circled{6}).
These artifacts include the data generated by the blockchain systems and the data output by the Gromit instances, e.g., monitoring statistics.
This data is parsed by the orchestrator (step~\circled{7}) and generates human-readable graphs.
Finally, the data is stored in a database (step~\circled{8}) and is ready for analysis by the user (step~\circled{9}).

\subsection{Integrating Blockchain Systems into Gromit}
To show practicality of Gromit, we have integrated seven prominent blockchain systems into it. The integration requires no change in the source code of the blockchain systems. This allows to benchmark the blockchain system as close to the deployed systems as possible. 

The design of Gromit is modular, and developers can implement \emph{modules} that enrich an experiment with more functionality, for example, bandwidth monitoring.
Integration of a blockchain system requires a developer to subclass the \texttt{BlockchainModule} and to implement the \texttt{init\_configuration}, \texttt{start\_validator}, \texttt{stop\_validator}, and \texttt{parse\_ledger} methods.
Gromit enables developers to specify custom network topologies to connect validators among each other.
We refer readers who wish to integrate a particular blockchain system to the Gromit documentation.

Besides supporting blockchain systems, Gromit has support for generic experiments with distributed systems. For example, we have used Gromit to conduct experiments with peer-to-peer protocols on custom infrastructure.

\subsection{Transaction Analysis}
A workload during a Gromit benchmark consists of transactions issued by one or more clients. All transactions can be submitted to one validator peer or can be evenly spread among the peers. 
In line with related work, we mainly gauge blockchain performance using two metrics, transaction throughput, and transaction latency. Specifically, we consider the peak transaction \emph{throughput}, which is the maximum rate at which the system can process transactions before getting congested. 
Second, we analyze the \emph{latency} of transactions, which is the time between submitting a transaction and its irreversible inclusion in the distributed ledger.

We store the timestamp at which a client has submitted the transaction to a validator to determine transaction latency.
However, determining the finalization time of transactions can be complicated since some blockchain systems do not expose granular, temporal information.
We use the following two approaches to determine the latency of individual transactions.
Our first approach is to inspect the resulting distributed ledger after all transactions have been issued (this logic should be part of the \texttt{parse\_ledger} method).
We then compute transaction latency based on timestamps included in ledger data elements (e.g., blocks). However, this approach is useful only if the blockchain system annotates data elements in the distributed ledger with a timestamp.
The second approach suggests clients periodically poll the blockchain system to determine if a transaction has been confirmed.


\section{Blockchain Consensus Model}
\label{sec:consensus_families}

Since the introduction of Bitcoin in 2008, there have been many proposals for new blockchains and consensus mechanisms.
Some proposals have materialized into operational systems, whereas other ones are only theoretically analyzed.
The proliferation of blockchain solutions makes it infeasible to conduct a benchmarking study with all available systems.

\begin{table*}
	\small
	\centering
	\begin{tabular}{ | p{3cm} | p{4.9cm} | p{3cm} | p{2.4cm} | p{2cm} |}
		\hline
		\textbf{Blockchain System} &
		\textbf{Consensus Model} 
		& \textbf{Consensus Principle}
		& \textbf{Validator Group Formation}
		& \textbf{First commit (year)}
		\\ \hline
		Ethereum (v1.9.24)~\cite{wood2014ethereum} & 
		Proof of Work (PoW) 
		& Resource-based lottery & Resource mining &  2013
		\\ \hline 
		Algorand (v2.3.0)~\cite{gilad2017algorand}  & Proof of Stake (PoS) & 
		Random selection of leaders & 
		Stake-based enrolment & 2019
		\\  \hline
		BitShares (v5.0.0)~\cite{schuh2017bitshares} & Delegated Proof of Stake (dPoS) & Rotating leader & Election by stakeholders & 2015
		\\ \hline
		Diem (v1.1.0)~\cite{libra2019} &
		Practical Byzantine Fault Tolerance (PBFT), based on HotStuff~\cite{yin2019hotstuff} 
		&  Leader-based & 
		Enrolment by an authority &  2019
		\\ \hline
		Stellar (v15.1.0)~\cite{lokhava2019fast} &
		Federated Byzantine Agreement (FBA)  & Quorum Intersection & User-defined quorums
		&  2014 
		\\ \hline
		Hyperledger Fabric (v1.4.9)~\cite{androulaki2018hyperledger} & 
		Crash-tolerant consensus (CFT), based on Raft~\cite{ongaro2014search} & 
		Leader-based & 
		Enrolment by an authority &  2016 
		\\ \hline
		Avalanche (v1.1.1)~\cite{rocket2019scalable} & 
		Meta-Stable Consensus (MSC) & Network subsampling & Stake-based enrolment &  2020\\ \hline
	\end{tabular}
	\caption{The seven selected blockchain systems and consensus models analysed in this work.} \label{tab:systems_comparison}
\end{table*}

Based on taxonomy of different consensus mechanisms used in blockchain systems~\cite{bano2019sok} we consider seven prominent consensus models and select a representative blockchain system for each consensus model.
Our selection process is based on the economic magnitude, adoption, maturity of the system, and the protocol's academic significance.
Table~\ref{tab:systems_comparison} lists the representative system for each considered consensus model. For each system, we also show the evaluated version of the software, the principle underpinning each consensus model, how the group of validators is formed, and the year in which the first software commit has been made (as an indicator of matureness).
We are aware of state-of-the-art blockchains that use techniques such as layer-one scaling to improve throughput, e.g., OmniLedger~\cite{kokoris2018omniledger} and RapidChain~\cite{zamani2018rapidchain}, or layer-two scaling solutions~\cite{gudgeon2020sok}. 
However, the primary focus of our benchmarking study is on deployed layer-one blockchain systems.
We now elaborate on each consensus model and refer the reader to \cite{bano2019sok} for a more extensive overview of blockchain consensus models.

\textbf{Proof-of-Work (PoW)} is the oldest consensus mechanisms designed explicitly for blockchain systems. 
PoW is used in blockchains such as Bitcoin~\cite{nakamoto2019bitcoin} and Ethereum~\cite{wood2014ethereum} and to date remains a standard approach to build open blockchain systems. PoW scales well in the number of participants: the Bitcoin network has over 11'000 operational miners.
However, PoW is known to consume large amounts of energy, and the achievable transaction throughput is theoretically limited to around 60 transactions per second~\cite{gervais2016security}.

\textbf{Proof-of-Stake (PoS)}, initially proposed by the Peercoin cryptocurrency~\cite{king2012peercoin}, is an alternative consensus mechanism that addresses the excessive resource usage by PoW. PoS is typically based on a random, periodic selection of a leader. This selection process is weighted by the participants' stake in the system, e.g., by the number of assets owned, or by the age of possessed assets. Algorand is one of the most prominent PoS-based blockchains and leverages Verifiable Random Functions (VRFs) for the leader election process~\cite{gilad2017algorand}.

\textbf{Delegated Proof-of-Stake (dPoS)} is a consensus mechanism where stakeholders elect a group of \emph{delegates} through voting.
Voting decision can, for example, be based on community engagement.
Since the set of validators participating in consensus remains relatively small compared to PoS-based solutions, dPoS-based consensus has in theory a better potential to scale.
BitShares is one of the most mature blockchains using dPoS consensus~\cite{schuh2017bitshares}.


\textbf{Practical Byzantine Fault Tolerance (PBFT)} is a consensus algorithm introduced in 1999~\cite{castro1999practical}.
PBFT is specifically designed for networks with static and pre-approved membership and has been revised for adoption in blockchain environments.
Recent advancements have resulted in HotStuff~\cite{yin2019hotstuff}, a consensus protocol based on PBFT that reduces communication overhead and increases throughput.
HotStuff is at the core of Diem, a permissioned distributed ledger maintained by a consortium led by Facebook~\cite{libra2019}. 

\textbf{Federated BFT (FBFT)} is a consensus model that distinguishes itself from the approaches mentioned earlier by having validators explicitly specifying trust relations.
Stellar is one of the first systems to adopt FBFT consensus~\cite{lokhava2019fast}.
The Stellar Consensus Protocol (SCP) leverages a federated voting approach in which each validator votes on statements while ensuring that no two members of an overlapping quorum can confirm contradicting statements.

\textbf{Crash-tolerant Consensus (CFT)} is a consensus approach widely used to achieve fault tolerance, e.g., by Apache Kafka~\cite{kafka}. Unlike PBFT,
CFT is not resistant against arbitrary (Byzantine) behaviour but can withstand crash-stop failures of participants.
Notable algorithms achieving CFT are Paxos~\cite{lamport2001paxos} and Raft~\cite{ongaro2014search}.
Hyperledger Fabric, one of the most prominent industrial blockchains, is currently using Raft~\cite{androulaki2018hyperledger}.

\textbf{Metastable Consensus (MSC)} is a family of consensus algorithms that leverage network subsampling techniques to determine the validity of a transaction.
The idea is to repeatedly sample random validators in the network and to steer correct nodes to a common decision.
Avalanche is one of the most mature  blockchain solutions to leverage MSC consensus and maintains a DAG data structure to store transactions~\cite{rocket2019scalable}.

\begin{figure*}[t]
 \centering
 \begin{subfigure}{0.8\textwidth}
 \centering
		\includegraphics[width=0.8\linewidth]{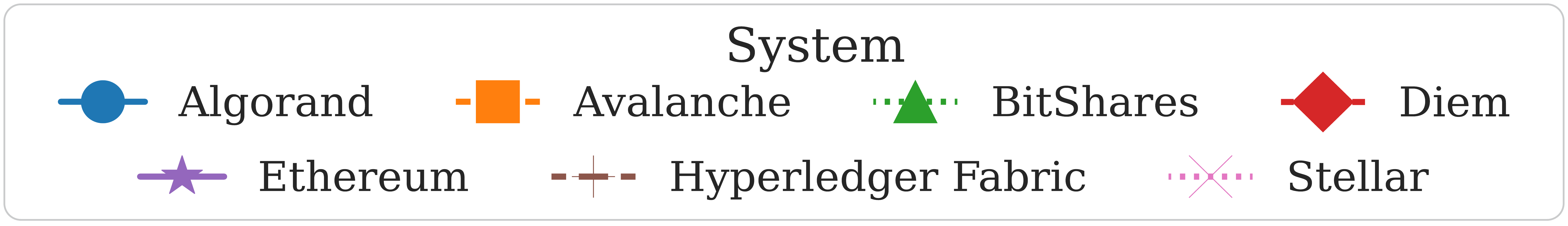}
	\end{subfigure}
	\centering
	\begin{subfigure}{.45\textwidth}
	 \vspace{0.2mm}
		\includegraphics[width=\linewidth]{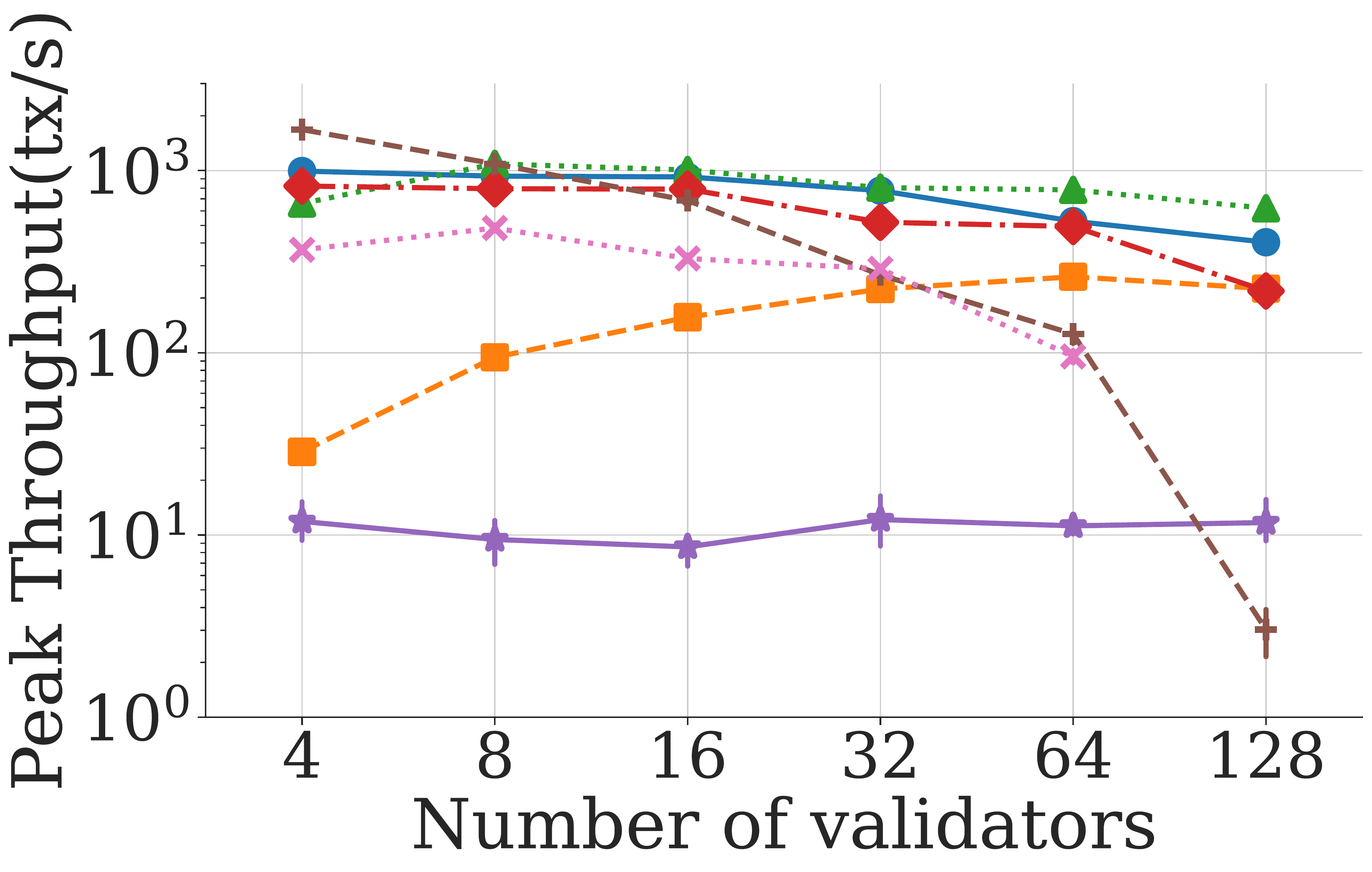}
		\caption{Peak transaction throughput}
		\label{fig:scalability_throughput}
	\end{subfigure}%
	\begin{subfigure}{.45\textwidth}
 	\vspace{-0.1mm}
		\includegraphics[width=\linewidth]{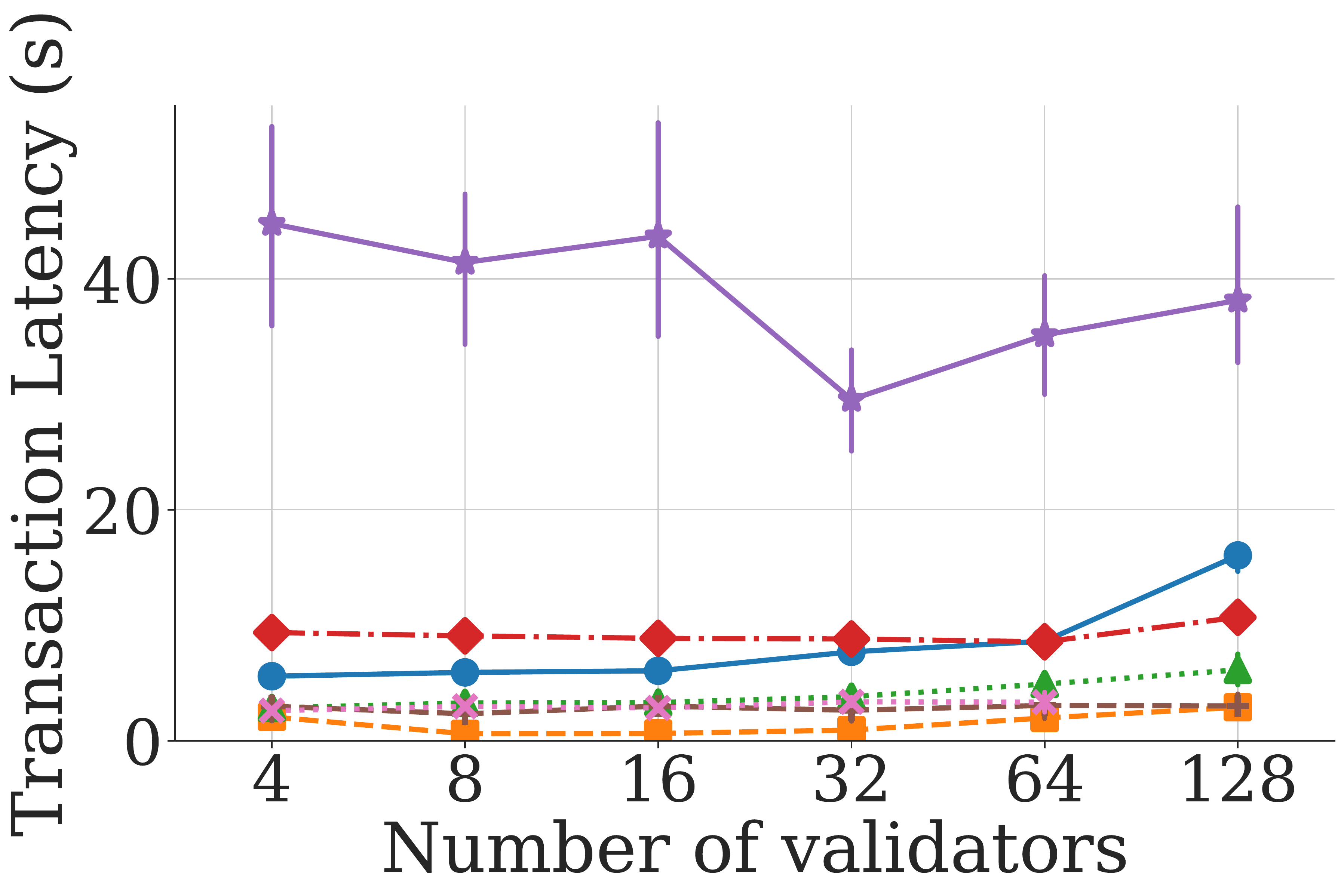}
		\caption{Transaction latency}
		\label{fig:scalability_latency}
	\end{subfigure}
	\caption{The peak throughput and transaction latencies of evaluated blockchain systems with the number of validators.}
	\label{fig:scalability}
\end{figure*}

\section{Blockchain Performance Evaluation}
\label{sec:evaluation}
We conduct a diverse set of experiments with Gromit to reveal the performance characteristics and limitations of seven blockchain solutions.
To the best of our knowledge, We are the first to perform a large-scale performance benchmark of blockchain solutions.


Our experiments answer the following questions:
How does the increase in the number of validators affect the system performance?
What is the peak performance of each system under a heavy system load?
What is the impact of a network delay on systems' performance?
How consistent are our performance results compared to previously reported values?


Throughout the section we use two variables for our experiments: $n$ indicates the number of validators and $\lambda$ signifies the transaction throughput. 

\subsection{Setup and Transaction Workload}

All experiments are conducted on four HPE DL385 Gen10 servers, located within the same data centre and inter-connected with 10GB Ethernet links.
Each server is equipped with 128 AMD EPYC 7452 CPUs, has 512GB of DDR4 memory, and runs Debian 10.
During our experiments, we deploy each blockchain system with its source code unmodified.
For each system, we use the default settings provided by the systems.
Each experiment starts with only the genesis block included in the blockchain.
We use a random network topology where each validator is connected to 10 other random validators.


We use Gromit to subject each system to a synthetic transaction workload, issued by up to 64 clients.
To ensure an equal load on each validator, a client submits transactions to the validator with ID $ i \equiv c\ (\mathrm{mod}\ n) $ where $ c $ is the ID of the client and $ n $ the total number of validators.
Each transaction is submitted to exactly one validator.
Transactions are submitted during a two-minute period, after which we wait an additional minute for all transaction to be finalized.


We use simple asset transfers as a performance baseline.
In our workload, a transaction issued by a client involves an asset transfer of a small, fixed amount to another account; the client counterparty is fixed throughout the experiment.
We ensure that each client has sufficient funds to spend during the experiment.
For Ethereum and Hyperledger Fabric, transactions involve the transfer of an ERC20 token.

\subsection{Determining Peak Transaction Throughput}
To determine peak transaction throughput, we gradually increase the system transaction rate in steps of 100 transactions per second (tx/s).
Based on the reported statistics by Gromit, we estimate the peak transaction load that each system is still able to process during a sustained period.
If the system has any unconfirmed transactions after our two-minute period, we consider the system as \enquote{saturated}.
We evaluate the peak throughput of each system with an increasing number of validators ($ n $).
We provide each system with an equal amount of resources and ensure that the resource usage of evaluated systems (CPU power, disk space, and memory) does not exceed the available resources.
Due to excessive resource usage of the Stellar software, we are only able to run Stellar with up to $64$ validators.
We run each experiment at least five times and average all results.
Appropriate graphs are annotated with 95\% confidence interval markers.
\\\
\\\
\textbf{Finding 1.}
\emph{Adding validators does not have a significant positive effect on the achievable peak transaction throughput of the evaluated systems.} \\

Figure~\ref{fig:scalability} shows the result of our scalability experiment as $ n $ grows, in terms of peak transaction throughput and transaction latency.
Figure~\ref{fig:scalability_throughput} shows the peak transaction throughput of evaluated systems (with a horizontal and vertical log-axis).
We notice that none of the evaluated systems can process over 1'000 tx/s with $ n = 128 $.
In general, the transaction throughput of most of the systems is capped between 500 and 1'500 tx/s.
Except for Ethereum, the peak transaction of all blockchains is \emph{decreasing} as $ n $ increases. Specifically, 
Hyperledger Fabric shows a severe degradation in performance when $ n > 8 $, and is just capable of processing 2 to 4 tx/s with $ n = 128 $.
We believe that this is caused by the underlying consensus model of Hyperledger Fabric, Raft, which does not scale well with the number of validators~\cite{ongaro2014search}. 
Of all systems, Ethereum has the lowest transaction throughput (around 10-20 tx/s), yet manages to keep stable throughput with the increase in the number of validators.
Peak throughput of Avalanche increases up to $n=64$, but then degrades for $n=128$.

\noindent \\ \textbf{Finding 2.}
\emph{For Avalanche, BitShares and Hyperledger Fabric, we observe significant discrepancies between the peak throughput found by us and previously reported values.} \\

\begin{figure*}[t]
	\centering
	\includegraphics[width=.8\linewidth]{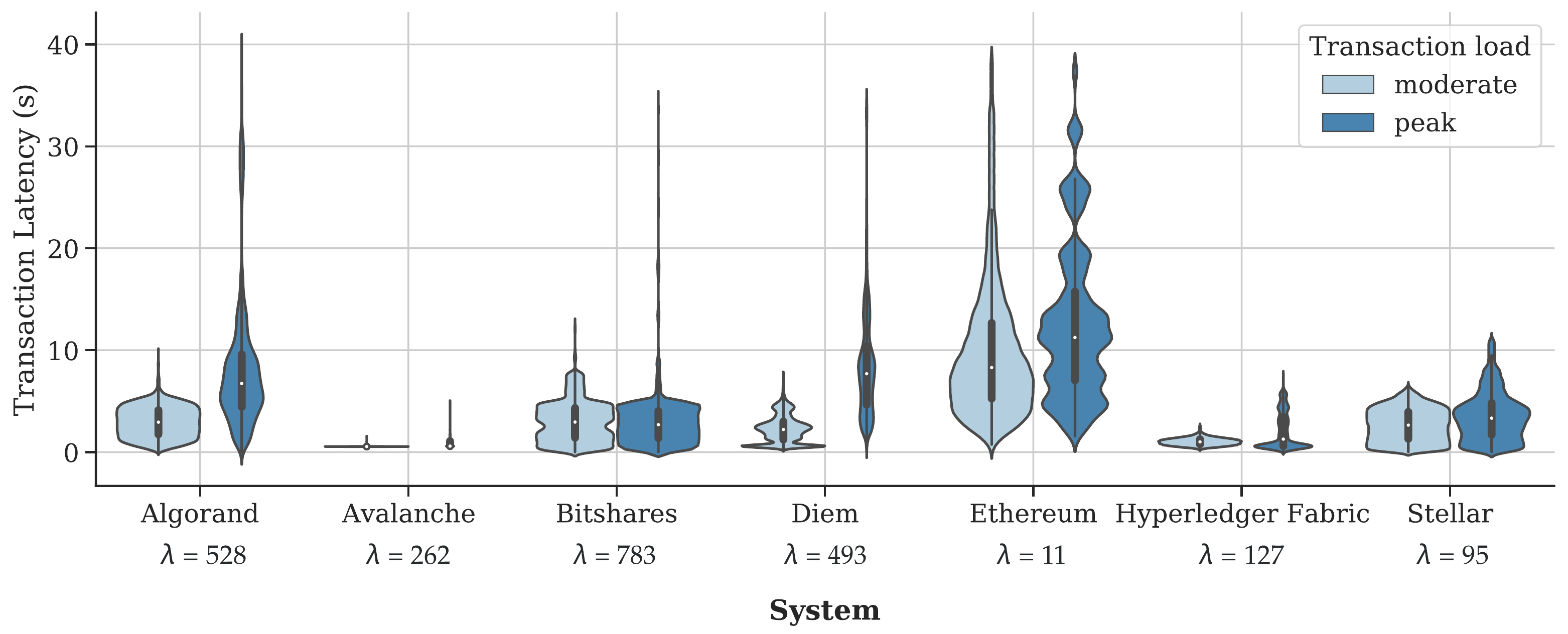}
	\caption{The distribution of transaction latencies for systems, under peak load ($ \lambda $ tx/s) and moderate load ($ \frac{\lambda}{2} $ tx/s) ($n = 64$).}
	\label{fig:latency_load}
\end{figure*}

The previous experiment estimates the peak throughput of blockchain systems, without modification to the source code and under default settings.
We now compare the performance results with values reported by other literature.
The performance, i.e. peak throughput, is typically determined through an evaluation by system developers themselves.
For Stellar, we could not find reliable benchmarking results, making this work the first benchmarking study of Stellar.
We are interested to see if there are significant inconsistencies with performance metrics reported by these studies.

Our performance results are comparable with results reported for Ethereum~(4-40 tx/s~\cite{pongnumkul2017performance}), Algorand~(880 tx/s~\cite{gilad2017algorand}) and Diem~(200-1'000 tx/s~\cite{zhang2019performance}). 
However, we notice that previously reported values for some systems are significantly higher than our findings, specifically for Avalanche~(7'000 tx/s, 26x higher~\cite{rocket2019scalable}), BitShares~(3'300 tx/s, 3x higher~\cite{bitsharesperformance}) and Hyperledger Fabric~(3'500 tx/s, 2 times higher~\cite{androulaki2018hyperledger}).
For each system, we now explain these inconsistencies with additional experiments and analysis.

\textbf{Avalanche.}
The relative low throughput of Avalanche surprises us and warrants further performance analysis.
Since we noticed that each validator node is fully utilizing a CPU core, even with $ n = 4 $ and 32 tx/s, we perform a CPU analysis of deployed validators using the \texttt{pprof} profiler.
We find that around 60\% of CPU time is spent on hash computations using the \texttt{argon2} algorithm~\cite{biryukov2016argon2}.
This CPU consumption originates from the API provided by Avalanche validators.
Specifically, each validator maintains a keystore with credentials that is managed by end users; interactions with that keystore, e.g., accessing a private key, requires the user to include the password hash in the request.
Consequently, many parallel requests to the API by clients cause severe performance degradations.

To analyse the impact of password hashing, we recompile Avalanche with this hash verification disabled and re-run our experiment.
We do not observe a significant increase in transaction throughput.
However, this reveals another performance bottleneck, originating from the verification of transactions, consuming around 90\% of CPU time.
Since Avalanche transactions are linked in a DAG structure, incoming transactions require the validation of parent transactions, which is a resource-intensive, recursive operation.
We believe this can be addressed with further engineering efforts, e.g., queueing the verification of transactions.



\textbf{BitShares.}
We further analyse the reported throughput of BitShares and found that the peak transaction throughput (3'300 tx/s) is not an accurate performance indicator since the sustained throughput throughout the experiment is only around 450 tx/s.
Specifically, the achievable throughput of the BitShares consensus algorithm seems to be predicated by the speed of the slowest consensus participants in terms of connectivity and CPU resources. When operating BitShares in a heterogeneous environment, this can result in significant deviations in transaction throughput.

\textbf{Hyperledger Fabric.}
We further analyse the results reported in the work of Androulaki et al.~\cite{androulaki2018hyperledger} and Blockbench~\cite{dinh2017blockbench}. 
We find that their work evaluates an early implementation of Hyperledger Fabric using a different consensus model (Zookeeper or PBFT). As such, these results are not directly comparable.

We also present the following two reasons to explain the discrepancies in reported and observed throughput numbers:
\begin{enumerate}
	\item \emph{Experimental Software vs Production}. Many of the throughput numbers reported by system developers are extracted using a premature or even incomplete software implementation.
	As such, we argue that the values found by our experiments are a more accurate reflection of the achievable throughput in a production environment.
	Additionally, we noticed that some solutions (Diem and Stellar) have built-in measures that artificially lower the achievable throughput, likely to ensure safety properties or to prevent attacks in a production environment. A similar insights were observed in~\cite{chacko2021my}. 
	\item \emph{Client-Validator Interaction}.
	In our experiments, clients submit transactions to the API exposed by the system, whereas other studies might directly inject transactions in the validator process.
	API-based interaction adds additional overhead as the request needs to be processed, and this approach therefore is likely to lower the peak throughput of the system.
	However, this approach resembles how users interact with validators when a blockchain system is Internet-deployed.
\end{enumerate}


\subsection{Transaction Latency}

\textbf{Finding 3.}
\emph{For all evaluated systems, the average transaction latency under peak load is largely independent of the number of validators.}\\

\begin{figure*}[t]
	\centering
	\includegraphics[width=0.9\linewidth]{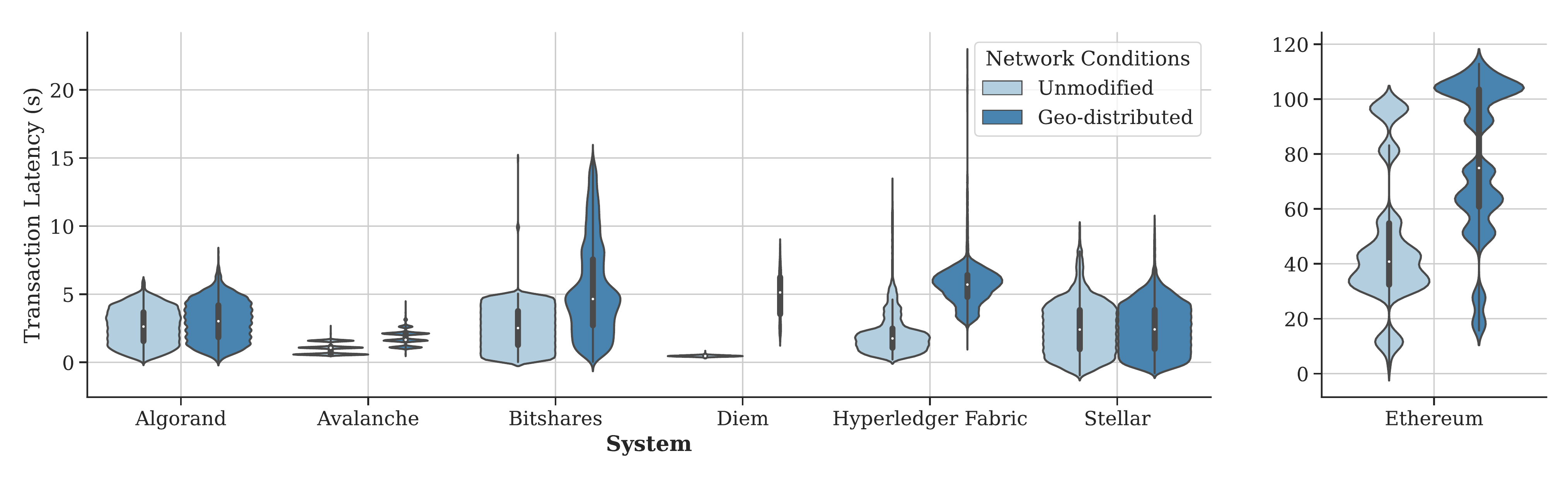}
	\caption{Transaction latencies without network modifications and in a geo-distributed setting ($ \lambda = 64 $ tx/s, $ n = 32 $). The transaction latencies of Ethereum are shown in the right plot.}
	\label{fig:latency_challege}
\end{figure*}

Figure~\ref{fig:scalability_latency} shows the average transaction latency under peak load, as $ n $ increases.
Except for Ethereum, the average transaction latency of evaluated systems is around or below ten seconds.
The variance between different runs is relatively low, except for Ethereum. 
In general, the average transaction latency increases as the number of validators grows. 
The consensus algorithm underlying BitShares, Algorand and Stellar progresses in five-second rounds, theoretically resulting in an average transaction latency of 2.5 seconds. 
We see that the transaction latency of Algorand increases from 5 seconds for $ n = 4 $ to 15 seconds with $ n = 128 $, suggesting that consensus rounds take longer to complete.
For BitShares and Stellar, this increase is less pronounced.
\\\\
\textbf{Finding 4.}
\emph{The variance of transaction latencies for Algorand and Diem increases significantly under peak load, compared to a moderate load. However, the transaction latencies of BitShares and Ethereum are largerly indifferent towards the system load.}\\

We visualize the distribution of transaction latencies for each system to explore further the effects of increasing the system load on the transaction latency.
We consider both peak and moderate loads, the latter being defined as half the determined peak load.
Figure~\ref{fig:latency_load} shows this distribution in a violin plot.
We observe that Algorand, BitShares, Hyperledger Fabric, and Stellar transaction latencies are roughly uniformly distributed under moderate load.
These systems adopt a round-based consensus approach, with a target of around five seconds for Algorand, BitShares, and Stellar, and one second for Hyperledger Fabric.
For these systems, most transactions are usually confirmed within the current or next consensus round relative to transaction submission.
The distribution of transaction latencies transforms as the system is subjected to a peak load.
Figure~\ref{fig:latency_load} shows that the finalization of Algorand transactions is being deferred to later rounds: 6\% of Algorand transactions have a transaction latency above 20 seconds.
This effect is less pronounced for BitShares and Stellar.

Increasing the system load impacts the latency distribution of Diem transactions.
Further investigation reveals that the round duration in Diem adjusts to the system load.
As more validators join the network and as more transactions are submitted to Diem validators, the round duration increases to ensure transactions can be processed on time.
Nonetheless, 30\% of all issued transactions in Diem are confirmed only after 10 seconds under peak load, whereas the system can handle all submitted transactions within 7 seconds under moderate load.

\subsection{Impact of Network Delays}
\textbf{Finding 5.}
\emph{Adding network delays has a minimal effect on the transaction latencies of Algorand and Stellar. However, Avalanche and Diem are extremely sensitive to network delays.}\\

Finally, we modify the network settings using the \texttt{netem} Linux kernel module\footnote{TC documentation: \url{https://www.linux.org/docs/man8/tc-netem.html}} and measure the impact of a network delay on the transaction latency for integrated blockchain systems.
This experiment reveals how different blockchain systems are sensitive to deployment in geo-distributed settings. To replicate realistic network delays, we extract ping statistics for 32 cities worldwide from WonderNetwork~\cite{WonderNetwork} and assign each validator to a city in a round-robin fashion.

Figure~\ref{fig:latency_challege} shows the distribution of transaction latencies of systems deployed in one data center with unmodified network settings and when deployed with geo-distributed settings.
For each experiment run, we use 32 validators and fix the transaction load to 64 tx/s, which all evaluated systems can handle without congestion.
Except for Algorand and Stellar, network delays visibly impact transaction latencies.

The effect is the most pronounced for the permissoned systems Diem and Hyperledger Fabric. This suggests that these systems can only operate in controlled/closed network environments, as any change in the network significantly affects the performance. We also observe a significant impact of network conditions on Avalance. This is due to the poll-based nature of metastable consensus. The poll rounds are visible in Figure~\ref{fig:latency_challege}.

For BitShares, we observe that validators occasionally fail to produce a block with higher network latency. For Diem, consensus progress stales a few seconds after starting the experiment, and rounds are timing out without confirming any transaction, violating system liveness.

\subsection{Network and CPU Utilization}

\textbf{Finding 6.}
\emph{Algorand, Stellar, Diem show high network utilization under idle load. Ethereum, Stellar and Diem consume significant CPU resources under idle load.}\\

\begin{figure*}[t]
    \begin{subfigure}{\textwidth}
		\includegraphics[width=\linewidth]{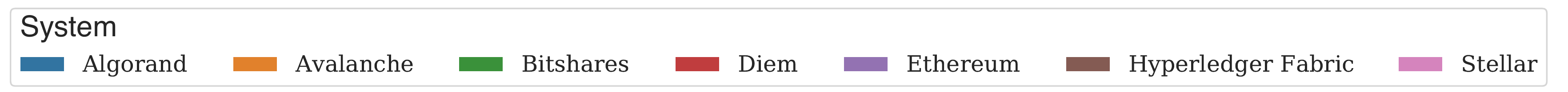}
	\end{subfigure}
	\centering
	\begin{subfigure}{.48\textwidth}
	    \vspace{0mm}
		\includegraphics[width=0.9\linewidth]{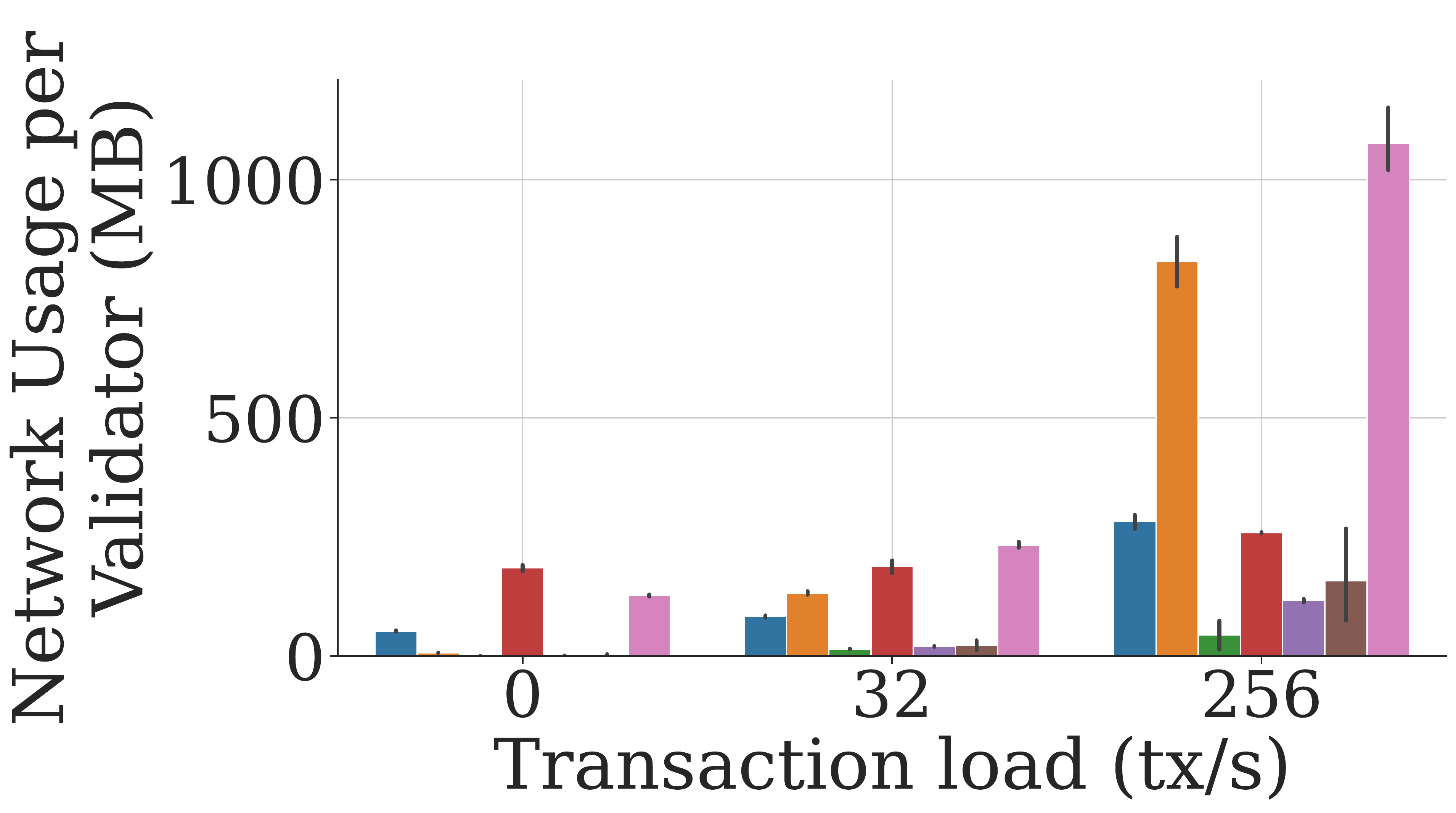}
		\caption{Network usage}
		\label{fig:bandwidth_load}
	\end{subfigure}%
	\begin{subfigure}{.48\textwidth}
	    \vspace{0mm}
		\includegraphics[width=0.8\linewidth]{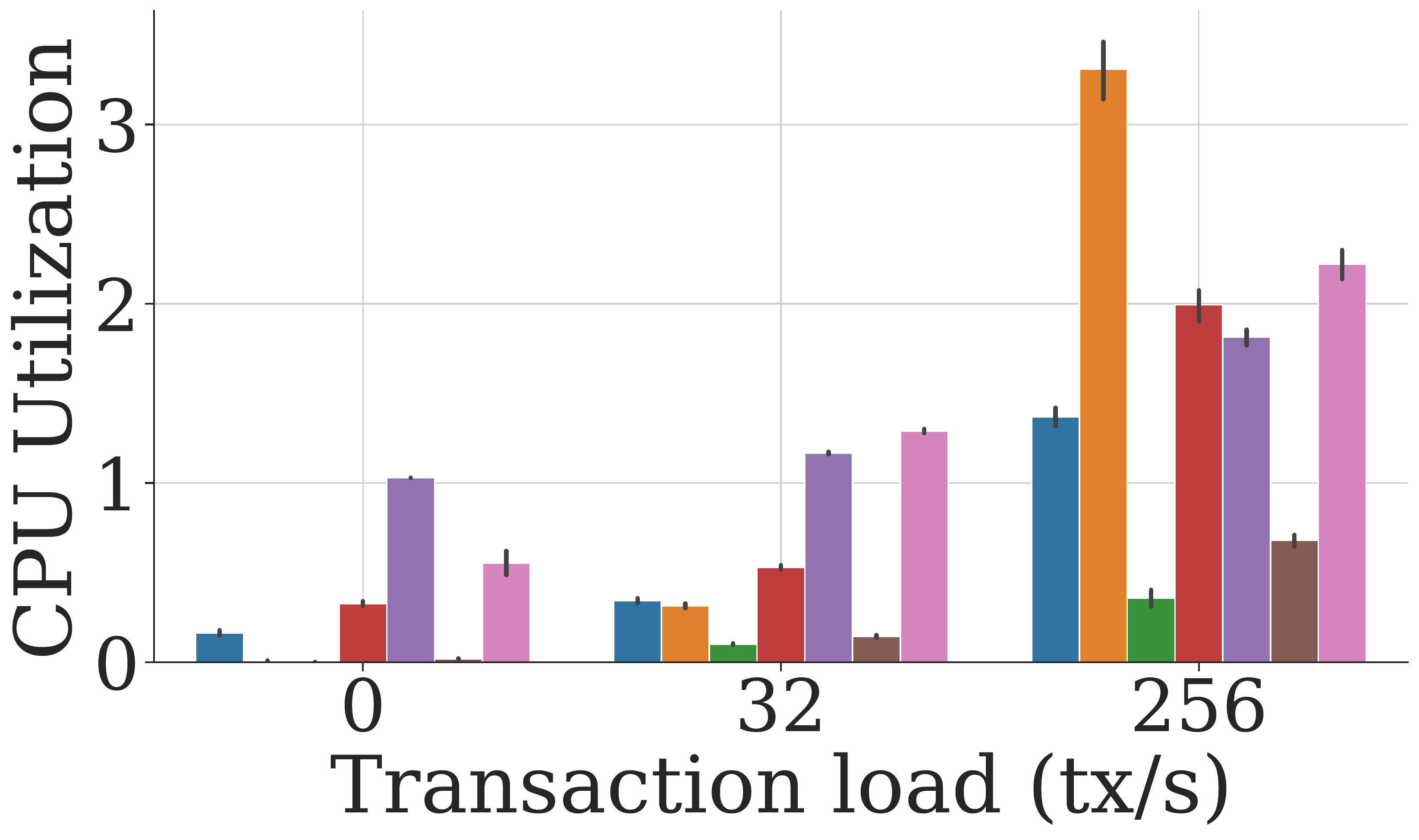}
		\caption{CPU utilization}
		\label{fig:utime_load}
	\end{subfigure}
	\caption{The resource utilization of the evaluated systems with increase in transaction load ($n = 32$).}
	\label{fig:resources_under_load}
\end{figure*}

\begin{figure*}[t]
	\centering
	\begin{subfigure}{.48\textwidth}
		\vspace{-3mm}
		\includegraphics[width=0.9\linewidth]{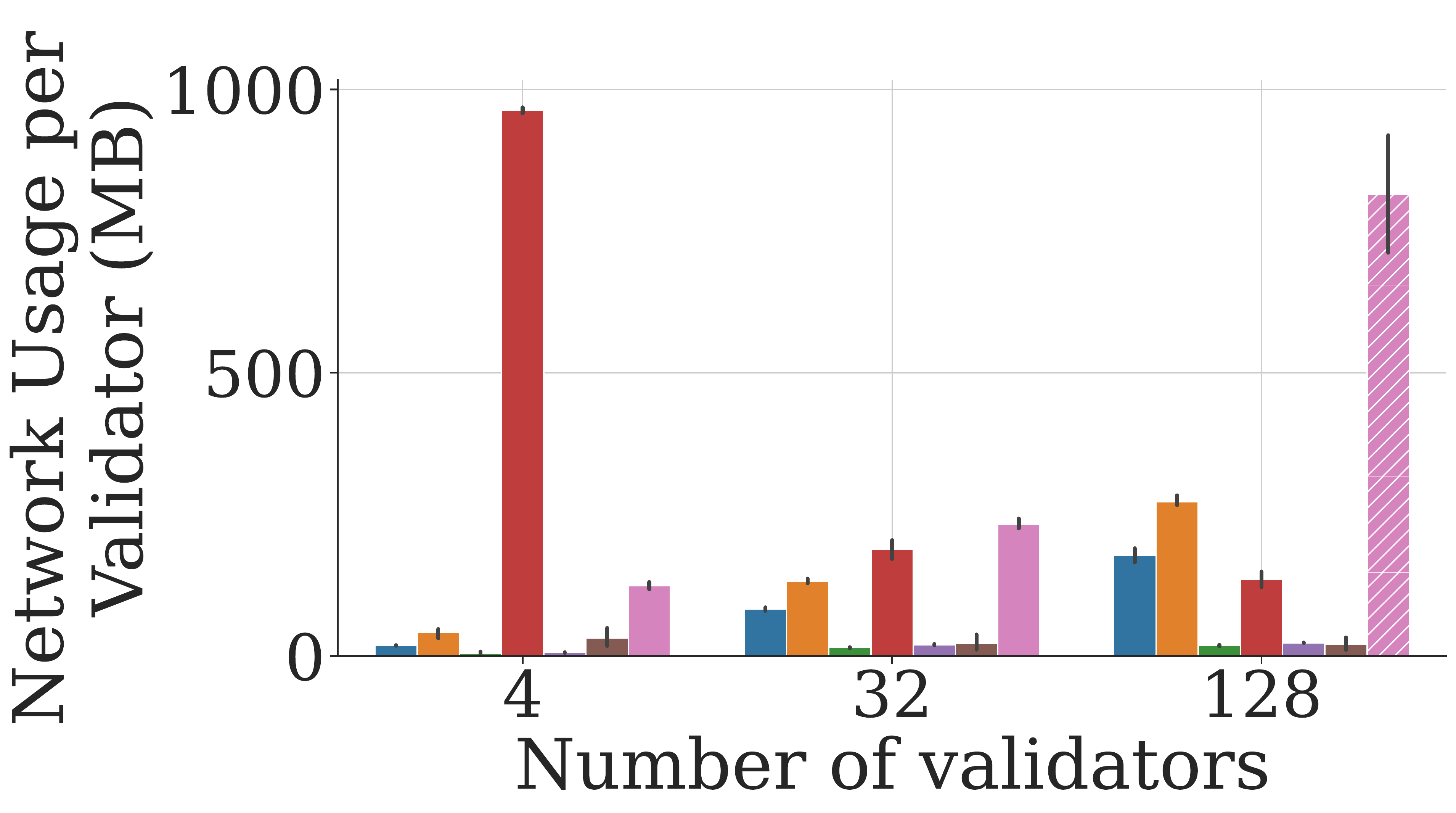}
		\caption{Network usage}
		\label{fig:scalability_resource_usage_bandwidth}
	\end{subfigure}%
	\begin{subfigure}{.48\textwidth}
		\vspace{-3mm}
		\includegraphics[width=0.9\linewidth]{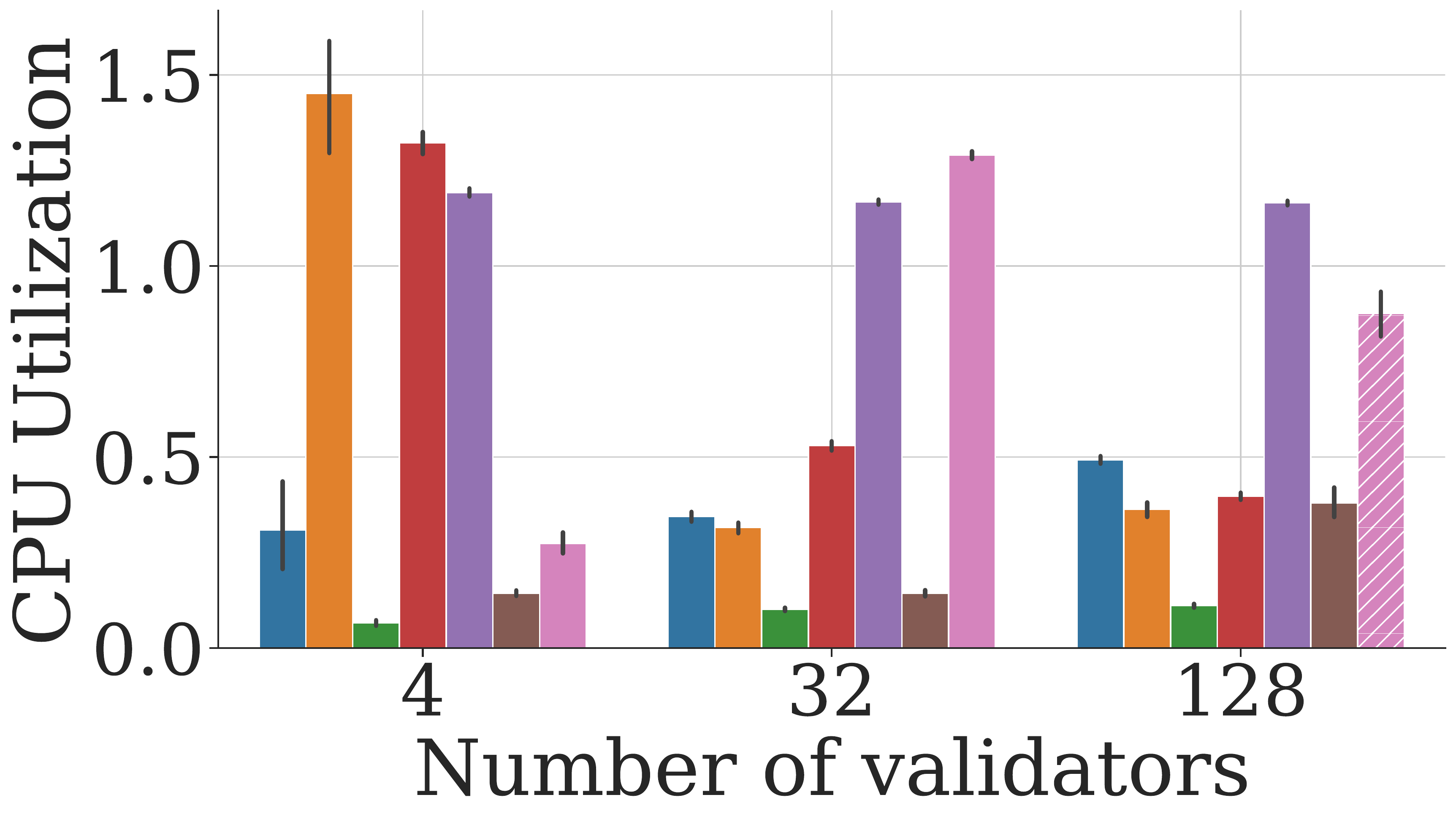}
		\caption{CPU utilization}
		\label{fig:scalability_resource_usage_cpu}
	\end{subfigure}
	\caption{The network and CPU usage for evaluated systems, when increasing the number of validators ($\lambda = 32 $ tx/s).}
	\label{fig:scalability_resource_usage}
\end{figure*}

We track the total network usage (inbound and outbound traffic) and average CPU utilization for each system while increasing $ \lambda $ and fixing $ n $ to 32.
To obtain insights into the system under idle load, we also run blockchain systems with $ \lambda = $ 0 tx/s.
Figure~\ref{fig:bandwidth_load} shows that the network usage per validator quickly grows for Avalanche and Stellar as the transaction load increases.
For $ \lambda = 256 $ tx/s, both Avalanche and Stellar use over 800 MB of network traffic per validator process.
BitShares is the most network-efficient, using only 80 MB per validator for $ \lambda = 256 $ tx/s.
We also observe that Algorand, Diem, and Stellar show 10x to 100x more bandwidth consumption under no transaction load compared to other systems.

Figure~\ref{fig:utime_load} shows the average CPU utilization when increasing $ \lambda $.
The mining process of Ethereum is continuously utilizing a single CPU.
Under $ \lambda = 256 $ tx/s, BitShares and Hyperledger Fabric are the most CPU-efficient compared to other systems. 
Our synthetic workload results are consistent with real-world observations of liveness issues for Avalanche and Stellar~\cite{ogrady_2021, mcsweeney_2021}.

Figure~\ref{fig:scalability_resource_usage_bandwidth} shows that Diem is using significant network resources for $ n = 4 $: 970 MB per validator process.
This number decreases quickly when $ n $ increases.
We explain this behavior by the self-adjusting round times of the Diem consensus mechanism: as $ n $ increases, rounds take longer to complete, lowering the bandwidth usage.
For Stellar, we see the opposite effect: network usage becomes significant for $ n = 128 $.
Although Stellar cannot process transactions for $ n=128 $, we report its resource usage nonetheless.
Inspection of Stellar logs reveals that validators lose track of consensus, clogging the network with resynchronization messages.

Figure~\ref{fig:scalability_resource_usage_cpu} shows how CPU utilization behaves when increasing $ n $.
BitShares is the most CPU-efficient for all evaluated values of $ n $.
The CPU load of validators decreases for Avalanche and Diem as $ n $ increases.
Since we fix the transaction load, adding more validators decreases the individual load.

\section{Related Work}
Blockchain benchmarking and its associated challenges has received attention from other researchers.
Fan et al. present an extensive survey outlining methods for evaluating blockchain performance~\cite{fan2020performance}.
The work of Wang and Ye describes benchmarking tools and consensus mechanisms and outlines techniques to improve the throughput of blockchains~\cite{wang2019performance}.
The authors of these studies summarize work on blockchain performance but do not conduct benchmarking studies themselves.


Popular blockchain benchmarking tools are Blockbench~\cite{dinh2017blockbench}, Hyperledger Caliper~\cite{caliper}, and DAGBench~\cite{dong2019dagbench}.
Blockbench, introduced in 2017, is the earliest benchmarking framework for blockchain and is specifically designed to evaluate permissioned blockchains~\cite{dinh2017blockbench}.
Blockbench measures the performance of components commonly found in blockchains, e.g., the transaction execution engine).
The authors of Blockbench evaluate the performance of Hyperledger Fabric, Ethereum, and Parity.
Hyperledger Caliper is a benchmarking tool for the performance evaluation of specific systems with a set of pre-defined use cases~\cite{caliper}.
Hyperledger Caliper primarily supports projects by the Hyperledger Foundation.
DAGBench is a benchmarking tool for DAG-based blockchains~\cite{dong2019dagbench}.
The authors evaluate three popular DAG-based blockchain implementations~\cite{pervez2018comparative}: IOTA, Nano, and Byteball.
However, DAGBench does not allow for a comparison of other blockchains. The authors of BCTMark~\cite{saingre2020bctmark}. present an Ethereum performance evaluation.




\section{Conclusion}
We have presented Gromit, a generic benchmarking framework for blockchain solutions.
By treating each blockchain system as a transaction fabric, our framework enables any blockchain system's integration, benchmarking, and performance analysis.
We leverage the functionalities of Gromit and conduct the largest blockchain benchmark to date, involving seven prominent blockchain systems.
Our main finding is that none of the evaluated solutions can handle beyond 1'000 transactions per second as the number of validators increases. Yet, they show relatively low transaction latencies on average.
We also find that synthetic workloads can provide accurate predictions of performance limitations, as our findings are consistent with real-world observations on performance incidents on Avalanche and Stellar networks.


\bibliographystyle{IEEEtran}
\bibliography{references}
\end{document}